\documentclass[twocolumn,aps,prb,showpacs,superscriptaddress,unsortedaddress]{revtex4}
\usepackage{graphicx}
\usepackage{epsf}

\begin{document}
\title{The nature of spectral gaps due to pair formation}

\author{T. Micklitz}
\affiliation{Materials Science Division, Argonne National Laboratory, Argonne, IL 60439}
\author{M. R. Norman}
\affiliation{Materials Science Division, Argonne National Laboratory, Argonne, IL 60439}

\begin{abstract}
Several phenomenological self-energies have been presented to describe
the pseudogap in cuprates.  Here, we offer a derivation of the self-energy in two
dimensions due to pair formation and compare it to photoemission data.
We then use our results to address several
questions of interest, including the existence of magneto-oscillations in
the presence of the pseudogap, and the two length scale nature of vortices in
underdoped cuprates.
\end{abstract}

\date{\today}
\maketitle


Various models for the self-energy have been presented to describe photoemission
spectra for the cuprate pseudogap phase.\cite{arc1}  The basic functional form is
\begin{equation}
\Sigma(k,\omega) = \frac{\Delta^2}{\omega - X_k + i\Gamma_0}
\end{equation}
where $\Delta$ is the energy gap~\cite{delta} and $\Gamma_0$ the broadening.
For the pairing scenario, $X_k = -\epsilon_k$, where $\epsilon_k$
is the single particle dispersion.  This Ansatz has a long history going back to the original
BCS theory,\cite{schrieffer}  where it implicitly describes broadening due to
impurities.\cite{ag}

Lee, Rice, and Anderson \cite{lee} were able to derive the same functional form 
for a one-dimensional density wave state, with $X_k = \epsilon_{k+Q}$
where $Q$ is the wavevector of the density wave.  In this one dimensional case, long
range order is not present ($\Delta^2 \equiv \left<\Delta^2\right>$).  The result was derived at lowest order restricting to static thermal fluctuations.  In this case,
$\Gamma_0$ is replaced by $\Gamma_2 = v_F/\xi$ where $v_F$ is the Fermi velocity
and $\xi$ the correlation length.  A similar derivation in two dimensions yields
instead \cite{maki}
\begin{equation}
-Im\Sigma(k,\omega) = \frac{\Delta^2}{\sqrt{(\omega+\epsilon_k)^2+\Gamma_2^2}}
\end{equation}

Eq.~1 was proposed some time ago to describe data in the pseudogap phase of
the cuprates.\cite{prb98,franz,levin}  In Ref.~\onlinecite{prb98}, it was motivated
by a `zero dimensional' approximation where the fermion dispersion is ignored 
(i.e., $\epsilon_{k-q} \sim \epsilon_k$) when
doing the momentum integration ($\Sigma \sim \int D G$ where $D$ is the boson
propagator and $G$ is the fermion Green`s function).
In this case, $\Gamma_0$ in Eq.~1 reduces to that
of time dependent Ginzburg-Landau theory, and should scale approximately as $T-T_c$
(as compared to the $\sqrt{T-T_c}$ behavior of $\Gamma_2$ in Eq.~2).
This was found to give a good account of the $T$ dependence of the photoemission data
above $T_c$ for underdoped cuprates at the antinodal points of the Brillouin zone 
(where the $d$-wave energy gap is largest).\cite{prb98}  It was claimed in this work that this functional
form could be motivated in higher dimensions as well, but as we show here, this is
dependent on the value of two physical parameters, $\Delta/T_c$ and $v_F/\xi_0T_c$.

Recently, Senthil and Lee \cite{senthil}  proposed a related Ansatz for the zero
temperature limit, which was motivated by a desire to address magneto-oscillation data
in the cuprates.  Their Ansatz, though, leads to three spectral peaks, as opposed to Eq.~1
that either yields two peaks (gapped case) or one peak (gapless case) depending on
the ratio $\Gamma_0/\Delta$.  Their result is similar to a related one derived for a
spin density wave by Kampf and Schrieffer.\cite{kampf} We note that both results seem to be
at variance with the expectation that the energy gap should be confined to the
ordered and `renormalized classical' phases, and therefore should not be present in
the zero temperature limit unless ordering is present.\cite{vilk97}

In this Rapid Communication, we provide a derivation of the fermion self-energy due to pairing
including both the static thermal fluctuations as in Refs.~\onlinecite{lee,maki} and
the dynamical fluctuations as in Ref.~\onlinecite{prb98}.   Above $T_c$, we find
a result in two dimensions
which contains aspects of both Eqs.~1 and 2, with the dynamical broadening ($\Gamma_0$)
dominating over the thermal broadening ($\Gamma_2$) if $v_F/\xi_0\Delta$ is small relative to
unity (this ratio is $\pi$ in BCS theory), where $\xi_0$ is the bare coherence length.
With a reasonable choice of parameters, we find that it quantitatively fits photoemission data for
underdoped cuprates.
At $T=0$, we find that for these same parameters, three spectral peaks are indeed present in
agreement with the work of Senthil and Lee, though for BCS parameters, only a single peak
occurs.


To lowest order, the electron self-energy is obtained by convolving the pair propagator with
the hole propagator:
\begin{equation}
\Sigma(k,\omega_n) = -T\sum_m\int \frac{d^dq}{(2\pi)^d}
D(q,\Omega_m) G_0(q-k,\Omega_m-\omega_n)
\end{equation}
where $D$ is the pair propagator and $G_0$ is the bare Green`s function
($G^{-1}_0=i\omega_n-\epsilon_k$), with the sum over boson Matsubara frequencies.
In the BCS approximation, $D$ is 
$\Delta^2 \delta(q)\delta(\Omega)$, immediately giving rise to Eq.~1 
with $\Gamma_0=0^+$.
In the absence of long range order,
\begin{equation}
D^{-1} = N_0 (x+\xi_0^2q^2+\alpha|\Omega_m|)
\end{equation}
with \cite{tinkham} $x \sim (T-T_c)/T_c$, $\alpha \sim \pi/(8T_c)$ and $\xi_0 \propto v_F/T_c$.
$N_0$ is the density of states per unit cell.
For $T > T_c$, the dominant contribution to the Matsubara sum comes from the branch 
cut of $D$ on the real axis.  This leads to ($\coth(\Omega/2T) ~ \sim 2T/\Omega$)
\begin{equation}
\Sigma(k,\omega_n) = \frac{T}{N_0}\int \frac{d^dq/(2\pi)^d}{x+\xi_0^2q^2}
\frac{1}{i\omega_n+i(x+\xi_0^2q^2)/\alpha+\epsilon_{q-k}}
\end{equation}
Evaluating (d=2 is assumed from here on), we find
\begin{equation}
\Sigma = \frac{-i\tilde{\Delta}^2}{\sqrt{(\omega+\epsilon_k)^2+\Gamma_2^2}}
\tan^{-1}\frac{\sqrt{(\omega+\epsilon_k)^2+\Gamma_2^2}}{-i(\omega+\epsilon_k)+\tilde{\Gamma}_0}
\end{equation}
with $\tilde{\Delta}^2 = \frac{T}{2\pi N_0 \xi_0^2}$,
$\Gamma_2 = v_F\sqrt{x}/\xi_0$ and
$\tilde{\Gamma}_0 = 2x/\alpha$.
Although this formula does a good job of reproducing the filling in of the pseudogap with temperature
seen by photoemission,
the $T$ dependence of the spectral gap magnitude is not properly reproduced -
in particular, the spectral peak position exceeds $\tilde{\Delta}$ in magnitude for a large range
of $T$.  This problem can be traced to the definition of $\tilde{\Delta}$ itself.
In Ref.~\onlinecite{prb98}, this difficulty was avoided in the derivation of Eq.~1 by
ignoring the $q$ dependence of the second term in Eq.~5.  By making this
approximation, this term could be
extracted outside the $q$ integral.  The $q$ integral then reduces to the definition of the
fluctuational gap, $\left<\Delta^2\right>$.
The issue, though, is that the static terms giving rise to $\Gamma_2$ are ignored in this
approximation.

These troubles ultimately stem from the fact that Eq.~4 is a low $q$, low $\Omega$ approximation
of the true pair propagator.
Use of Eq.~4, though, closely matches the exact result if the thermal approximation ($\coth$ replaced
by $2T/\Omega$) is used to cut-off the $\Omega$ integration, and the $q$ integral is cut-off
at $1/\xi_0$.\cite{LT}  Evaluating Eq.~5 with the cut-off gives
\begin{equation}
\Sigma = -\frac{T}{4\pi N_0\xi_0^2}\frac{1}{\sqrt{c}}\ln\frac{2\sqrt{c}\sqrt{a+\frac{b}{x+1}+\frac{c}{(x+1)^2}}
+\frac{2c}{x+1}+b}{2\sqrt{c}\sqrt{a+\frac{b}{x}+\frac{c}{x^2}}
+\frac{2c}{x}+b}
\end{equation}
where $a=-1/\alpha^2$, $b=-v_F^2/\xi_0^2+2i(\omega+\epsilon_k)/\alpha$
and $c=(\omega+\epsilon_k)^2+xv_F^2/\xi_0^2$.  At high frequencies, this reduces to
\begin{equation}
\Sigma_{high} = \frac{T}{4\pi N_0\xi_0^2\omega}\ln\frac{x+1}{x}
\end{equation}
Noting that in this approximation, the fluctuational gap is
\begin{equation}
\left<\Delta^2\right> =  \frac{T}{4\pi N_0 \xi_0^2} \ln\frac{x+1}{x}
\end{equation}
we now find the proper high frequency behavior of the self-energy, $\left<\Delta^2\right>/\omega$.

Formally, $\left<\Delta^2\right>$ has a singular temperature dependence, but for purposes here,
we will simply set its value to experiment, noting that  photoemission spectra indicate
no temperature dependence of $\Delta$ at the antinode.\cite{prb98}
In Fig.~1a, the real and imaginary values of the self-energy from Eq.~7 versus $\omega$ are shown
for $x=0.1$,
and in Fig.~1b the half width half maximum of the imaginary part 
is shown versus $x$.  The parameters used were $\Delta/T_c=4$ and
$v_F/\xi_0T_c=1$.  These values were chosen so as to give a good account of the experimental
half width versus $x$ extracted from fitting photoemission data on underdoped cuprates using 
Eq.~1.\cite{prb98}
We note that the $\Delta/T_c$ ratio of 4 (as compared to the BCS value of 1.76) is a typical value 
observed in cuprates.  The value of $v_F/\xi_0T_c$ of 1 (as  compared
to the BCS value of $1.76\pi$) acts to emphasize the dynamic broadening ($\tilde{\Gamma}_0$) over
the static broadening ($\Gamma_2$).  The ratio of these two values is only 1/4 (compared to the
BCS value of $\pi$) and will have further consequences as discussed below.
As mentioned before, the resulting spectral functions have either two peaks or one peak
depending on the magnitude of the half width relative to $\Delta$,
with examples shown in Fig.~2.  This crossover over from gapped to gapless behavior occurs
when the ratio of the half width to $\Delta$ is about $\sqrt{2}$, the same as from Eq.~1.


\begin{figure}
\centerline{\includegraphics[width=3.4in]{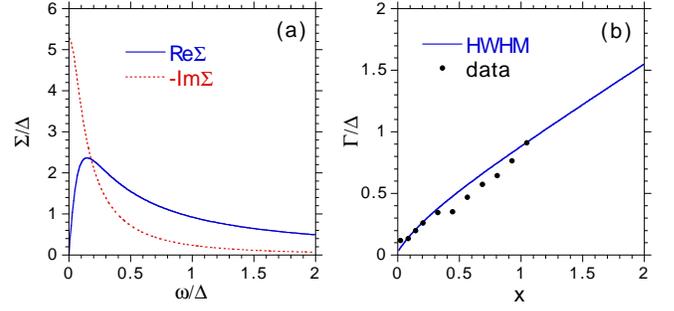}}
\caption{(Color online) (a) Self-energy from Eq.~7.  Parameters are $x \equiv (T-T_c)/T_c$=0.1,
$\Delta/T_c=4$ and $v_F/\xi_0T_c=1$.  (b) Half width of Im$\Sigma$, denoted as $\Gamma$,
versus $x$ compared to the data of Ref.~\onlinecite{prb98}.}
\label{fig1}
\end{figure}

\begin{figure}
\centerline{\includegraphics[width=1.7in]{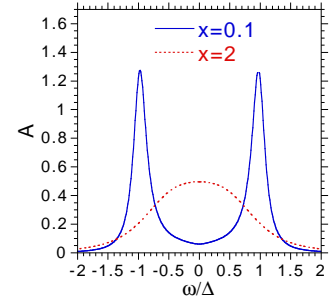}}
\caption{(Color online) Spectral functions ($\epsilon_k=0$) using the same parameters as Fig.~1
for $x=0.1$ and $x=2$.}
\label{fig2}
\end{figure}

In the zero temperature limit, the imaginary part of the self-energy is given by
\begin{eqnarray}
-Im \Sigma(k,\omega) & = & \int \frac{d^dq}{(2\pi)^d}  \int \frac{d\Omega}{2\pi}
(sgn(\Omega) - sgn(\Omega-\omega )) \nonumber \\
& & ImG(\omega-\Omega,k-q) ImD(\Omega,q)
\end{eqnarray}
Without cut-offs, this integral is
\begin{equation}
-Im\Sigma = \frac{1}{4\pi N_0 \xi_0^2} Im \ln
\frac{\frac{\omega+\epsilon_k}{\Gamma_2}-\frac{i\Gamma_2}{\tilde{\Gamma}_0} +
\sqrt{1+\frac{(\omega+\epsilon_k)^2}{\Gamma_2^2}}}
{\frac{\epsilon_k}{\Gamma_2}-\frac{i\Gamma_2}{\tilde{\Gamma}_0} +
\sqrt{1+\frac{\epsilon_k^2}{\Gamma_2^2}+\frac{2i\omega}{\tilde{\Gamma}_0}}}
\end{equation}
where $x$ in Eq.~4 is now a tuning parameter besides temperature
- magnetic field, etc., i.e., $(H-H_{c2})/H_{c2}$ - with $x=0$
corresponding to the quantum critical point where long range order appears.\cite{foot1}
The result is that -Im$\Sigma$ always grows with frequency, saturating to a constant as
$\omega \rightarrow \infty$.   The real part of the self-energy can be obtained by numerical 
Kramers-Kronig,\cite{KK} and it is found that the resulting spectral function is gapless.
The reason is that formally, the integrals defining $\left<\Delta^2\right>$ are divergent,
so cut-offs must be invoked, this time not only in momentum, but also in frequency as well.
We choose to
cut-off the $q$ integral at $1/\xi_0$ and the $\Omega$ integral at $1/\alpha$.
Reevaluating, we find
\begin{eqnarray}
-Im\Sigma & = & \frac{1}{2\pi^2N_0\xi_0^2}
Im \int_{q_1}^{q_2} dq_x [c(x+1) \tan^{-1}c(x+1) \nonumber \\
& & - c(x) \tan^{-1}c(x)]
\end{eqnarray}
where $c(y)^{-1} = \sqrt{y+q_x^2+i \alpha (\omega - v_Fq_x/\xi_0)}$,
$q_1=\max(0,(\omega-1/\alpha)\xi_0/v_F)$ and $q_2=\min(\omega\xi_0/v_F,1)$
with $q$ now expressed in units of $1/\xi_0$.  Similarly, we find that
\begin{eqnarray}
\left<\Delta^2\right> & = & \frac{1}{8\pi^2 N_0 \xi_0^2 \alpha} [(x+1)\ln\frac{(x+1)^2+1}{(x+1)^2}-x\ln\frac{x^2+1}{x^2} \nonumber \\
& & +2\tan^{-1}(x+1)-2\tan^{-1}(x)]
\end{eqnarray}
$\left<\Delta^2\right>$ is then used to set the prefactor in Eq.~12.
We can now evaluate -Im$\Sigma$ by doing one numerical
integration.  We note that in this approximation, -Im$\Sigma$ vanishes beyond
a frequency $\omega_c = 1/\alpha + v_F/\xi_0$ due to the cut-off in $\Omega$.
In fact, we note that the various cut-offs define two other
frequency scales as well, $\omega_1 = 1/\alpha$ and $\omega_2 = v_F/\xi_0$, with $\omega_c$
being their sum.  $\omega_1$ is associated with the dynamic part of the
pair propagator, and $\omega_2$ with the static part.

In Fig.~3a, we plot the self-energy from Eq.~12, and in Fig.~3b the resulting
spectral function, for the same parameters as in Fig.~1a.  One clearly see the existence of
three spectral peaks.  We can contrast this behavior with that in Fig.~4, where we show the
same as Fig.~3, but now for BCS parameters.  In the latter case, the asymptotics of the self-energy
sets in at a frequency beyond $\Delta$, and therefore no spectral gap emerges.
Similar results are obtained if one replaces the propagator in Eq.~4 by
that in a magnetic field in the lowest Landau level approximation.

\begin{figure}
\centerline{\includegraphics[width=3.4in]{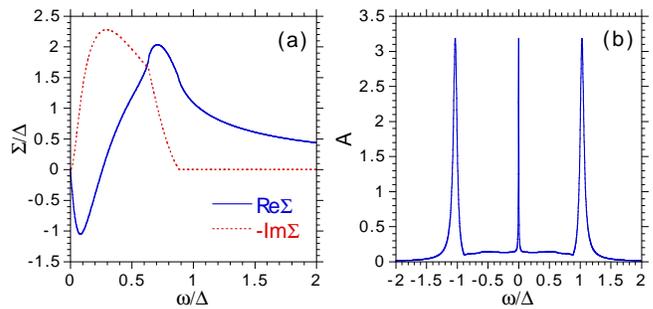}}
\caption{(Color online) (a) Self-energy from Eq.~12.  Parameters are $x=0.1$,
$\Delta/T_c=4$ and $v_F/\xi_0T_c=1$.
(b) Spectral function ($\epsilon_k=0$), where a constant 0.1$\Delta$
has been added to -Im$\Sigma$ so as to resolve the delta functions.}
\label{fig3}
\end{figure}

\begin{figure}
\centerline{\includegraphics[width=3.4in]{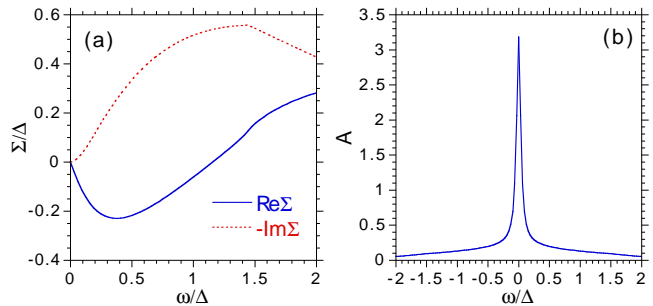}}
\caption{(Color online) Same as Fig.~3, but for
$\Delta/T_c=1.76$ and $v_F/\xi_0T_c=1.76\pi$.}
\label{fig4}
\end{figure}

Our $T$=0 results can be compared to the recent work of Senthil and Lee,\cite{senthil}
where a separable approximation
for the propagator was used.  In their work, a propagating form was considered
\begin{equation}
ImD = \frac{\Delta^2\pi^2\xi^{-1}}{(q^2+\xi^{-2})^{3/2}}
(\delta(\Gamma - \Omega)-\delta (\Gamma + \Omega))
\end{equation}
The resulting self-energy at $T=0$ is equivalent to that for electrons coupled to an Einstein mode 
with frequency $\Gamma$.\cite{schrieffer}  That is ($\omega > 0$)
\begin{equation}
-Im\Sigma = \frac{v_F}{2\xi}\frac{\Delta^2\Theta(\omega-\Gamma)}{(\omega+\epsilon_k-\Gamma)^2+v^2_F\xi^{-2}}
\end{equation}
where $\Theta$ is the step function.
This has a gap between $-\Gamma$
and $+\Gamma$ (with the real part of the self-energy diverging logarithmically at $\pm \Gamma$).
As a consequence, the spectral function consists of incoherent peaks
at $|\omega| > \Gamma$, and a quasiparticle pole within this gap.

A similar result occurs if one assumes a diffusive behavior which is more appropriate
for the disordered phase
\begin{equation}
Im D = -\frac{2\Delta^2\pi\xi^{-1}}{(q^2+\xi^{-2})^{3/2}}\frac{\Omega}{\Gamma^2 + \Omega^2}
\end{equation}
The resulting self-energy at $T=0$ is ($\epsilon_k=0$)
\begin{equation}
-Im \Sigma =  \frac{\Gamma\Delta^2}{\pi (4\Gamma^2+\omega^2)}\left(\frac{\omega}{\Gamma}
\tan^{-1}\left(\frac{\omega}{\Gamma}\right)+\ln\left(1+\frac{\omega^2}{\Gamma^2}\right)\right)
\end{equation}
where we have used that $\Gamma = v_F/\xi$.
This functional form (Fig.~5a) also leads to a spectral function with three peaks (Fig.~5b).

\begin{figure}
\centerline{\includegraphics[width=3.4in]{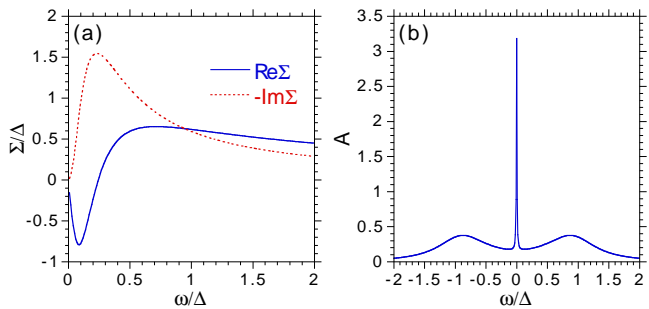}}
\caption{(Color online) (a) Self-energy from Eq.~17, with $\Gamma=0.1\Delta$.
(b) Spectral function ($\epsilon_k=0$), where a constant 0.1$\Delta$
has been added to -Im$\Sigma$ so as to resolve the quasiparticle pole.}
\label{fig5}
\end{figure}

It is interesting to note that in the Senthil and Lee formalism, the only energy scale is $\Gamma$,
and therefore a spectral gap occurs as long as the ratio of $\Gamma$ to $\Delta$ is not too large.
One reason for the difference from our work is that in their separable approximation,
$\Gamma$ is independent
of $q$, whereas from Eq.~4, one finds that the relaxational rate is strongly $q$ dependent,
that is $\Gamma_q = \alpha^{-1}(x+\xi_0^2q^2)$.\cite{tinkham2}  We also note that
formally, the $\Omega$ integral of Eq.~16 is logarithmically divergent when used to define
$\left<\Delta^2\right>$, but when calculating the self-energy, this is compensated for by the 
convergence of the $q$ integral in this
separable approximation.  That is, Eq.~17 is well behaved without the need to explicitly invoke
cut-offs.



We now turn to the question of the electron pockets observed by quantum oscillation
experiments.\cite{taillefer}  How can such pockets survive in the presence of a large
pseudogap, since  these electron pockets should originate in the antinodal regions of
the zone?\cite{leerpp}  As Senthil and Lee point out,~\cite{senthil} as one indeed finds a central
peak inside the gap in the low temperature limit, the existence of magneto-oscillations
is not a surprise (though in our case, we find a spectral gap only if the asymptotics
of the self-energy sets in below $\Delta$).  More generally, quantum oscillations
are seen in type II superconductors, sometimes for fields much less than $H_{c2}$.
At a semiclassical
level, this can be understood since the expectation value of the superconducting
order parameter averages to zero over a cyclotron orbit due to phase winding around the
vortices.  As a consequence, type II superconductors are gapless at high magnetic fields,
with the energy gap causing a broadening of the Landau levels.  Quantum mechanical
simulations have demonstrated the evolution of the low energy vortex core bound states into
Landau levels as the field is increased,\cite{nma} and similar calculations have been used to
address the quantum oscillation data in the cuprates.\cite{chen}  Extension of these methodologies 
to a potential vortex liquid phase above the resistive $H_{c2}$ would be illuminating.
We remark that the gapless peak in our work (and Senthil and Lee's) traces out a large 
Fermi surface, and therefore density
wave formation would have to be invoked to explain the small electron pockets that are
actually observed.\cite{millis}

We note that for the parameters in Figs.~1-3,
the value of $v_F/\pi\Delta$ is smaller than $\xi_0$ by a factor of 4$\pi$.  If we identify the former
with the size of the vortex core and assume a typical value of 30 $\AA$, then the
latter is approximately 400 $\AA$.  Such a long length has been identified from terahertz
conductivity measurements,\cite{orenstein} and implies a large `halo' which exists around the
vortex cores, leading to the concept of cheap, fast vortices, with the resistive $H_{c2}$ where
these halos overlap.\cite{senthil,rmp}  Therefore, a large $\Delta/T_c$ ratio and a small
$v_F/\xi_0\Delta$ ratio are conducive to obtain an extended regime above $T_c$ and $H_{c2}$
where an energy gap exists without long range order, a regime
that should be characterized by fluctuating vortices.
 
Work was supported by the U.S. DOE, Office of Science, under Contract 
No.~DE-AC02-06CH11357.  We thank Mohit Randeria, Todadri Senthil and Patrick Lee for discussions.


\begin{thebibliography}{99}
\bibitem{delta} $\Delta$ implicitly depends on $k$, i.e. $\Delta_k$.
This is allowed under the assumption that the pair wavefunction factorizes as a product
of relative and center of mass degrees of freedom.
\bibitem{foot1}
We assume that the inclusion of pair breaking (represented by $x$) leads to a functional
form for $D$ equivalent to Eq.~4.
\bibitem{KK}
The Kramers-Kronig is performed by linearizing Im$\Sigma$ over small intervals, each of
which is then analytically integrated and summed.
\end{thebibliography}
\end{document}